\documentclass[a4paper]{spie}  

\usepackage[]{graphicx}
\usepackage{aas_macros}
\usepackage[]{subfigure}

\title{The Gas Pixel Detector as an X-ray photoelectric polarimeter with a large field of view} 

\author{Fabio~Muleri\supit{a,b}, Paolo~Soffitta\supit{a}, Ronaldo~Bellazzini\supit{c},
Alessandro~Brez\supit{c}, Enrico~Costa\supit{a}, Sergio~Fabiani\supit{a}, Massimo~Frutti\supit{a},
Massimo~Minuti\supit{c}, Maria~Barbara~Negri\supit{d}, Michele~Pinchera\supit{c},
Alda~Rubini\supit{a}, Gloria~Spandre\supit{c}
\skiplinehalf
\supit{a} Istituto di Astrofisica Spaziale e Fisica Cosmica, Via del Fosso del Cavaliere 100,
I-00133 Roma, Italy;
\\
\supit{b} Universit\`{a} di Roma Tor Vergata, Dipartimento di Fisica, via della Ricerca Scientifica
1, 00133 Roma, Italy
\\
\supit{c} Istituto Nazionale di Fisica Nucleare, Largo B. Pontecorvo 3, I-56127 Pisa,  Italy
\\
\supit{d} ASI, Agenzia Spaziale Italiana, Viale Liegi 26, I-00198 Roma, Italy
}

\authorinfo{Further author information: (Send correspondence to Fabio Muleri)
\\Fabio Muleri: E-mail: Fabio.Muleri@iasf-roma.inaf.it,
Telephone: +39-0649934565}

\begin{document} 
\maketitle 

\begin{abstract}
The Gas Pixel Detector (GPD) is a new generation device which, thanks to its 50~$\mu$m pixels, is
capable of imaging the photoelectrons tracks produced by photoelectric absorption in a gas.
Since the direction of emission of the photoelectrons is strongly correlated with the direction of
polarization of the absorbed photons, this device has been proposed as a polarimeter for the study
of astrophysical sources, with a sensitivity far higher than the instruments flown to date.
The GPD has been always regarded as a focal plane instrument and then it has been proposed to be
included on the next generation space-borne missions together with a grazing incidence optics.
Instead in this paper we explore the feasibility of a new kind of application of the GPD and of the
photoelectric polarimeters in general, i.e. an instrument with a large field of view. By means of
an analytical treatment and measurements, we verify if it is possible to preserve the sensitivity
to the polarization for inclined beams, opening the way for the measurement of X-ray polarization
for transient astrophysical sources. While severe systematic effects arise for inclination greater
than about 20~degrees, methods and algorithms to control them are discussed.
\end{abstract}

\keywords{X-ray polarimetry, photoelectric absorption, large field of view instruments}

\section{Introduction}

Today polarimetry is the sub-topic of X-Ray Astronomy for which we have the maximum gap between the
expectations deriving from theoretical analysis and the achievements deriving from experiments
\cite{Meszaros1988, Weisskopf2006}. Indeed this branch of astronomy has suffered limitations due to
the lack of sensitivity of the previously flown X-ray polarimeters, based on the classical
techniques of measurement, namely the Bragg diffraction at nearly 45 degrees and the Thomson
scattering at nearly 90 degrees.

The development of new devices which can exploit the photoelectric effect to derive the
polarization of the absorbed photons has renewed the interest in this branch of X-ray astronomy.
Their operation is based on the reconstruction of the direction of emission of photoelectrons by
means of pairs created by ionization in a gas. Since photoelectric absorption is intrinsically
wide-band with respect to the Bragg diffraction and can be exploited at lower energies with respect
to the Thomson scattering, this kind of devices are characterized by a far higher sensitivity than
the X-ray polarimeters built so far.

The detection of polarization requires a large number of photons (10$^5$ - 10$^6$) to
reach minimum detectable polarizations of a few percent, i.e. in the range of interest for
astrophysical sources. To reach a sufficient collecting area to perform measurements in a reasonable
observing time, these devices can be efficiently employed with X-ray optics, and many proposals have
been presented\cite{Bellazzini2006b, Costa2006, Costa2007}. Two profiles of missions are emerging:
\begin{itemize}
\item a pathfinder mission, with a medium X-ray optics, which could perform the study of galactic
and bright extragalactic sources in a few days;
\item a large observatory, following the pathfinder mission, which could be dedicated to the study
of the faint extragalactic sources with a large area.
\end{itemize}
In both cases, the detector is aligned with the optics, namely the photons are incident
nearly orthogonally with respect to the instrument. Then, the symmetry around the optical axis 
assures that systematic effects are negligible, at least at first order.

Despite this essential employment of photoelectric polarimeters has not been realized yet, new
possibilities are already emerging. Indeed it has been proposed to employ this kind of polarimeters
to build small and cheap instruments with large field of view to perform polarimetry
of transient and bright sources like Gamma Ray Burst\cite{Hill2007}. 

In this paper, we analyze the possible use of the Gas Pixel Detector, one of the most advanced
project in the field of photoelectric polarimetry, as a large field of view instrument by means of
theoretical analysis and measurements. We will explore the systematic effects arising when
polarized and unpolarized photons are incident at large angle with respect to the perpendicular to
the instrument, and will present some basic tools to disentangle the systematics from the actual
polarized response. The bases of this analysis were already presented by Muleri in
2005\cite{Muleri2005}.

\section{The Gas Pixel Detector}

\begin{figure}[htbp]
\begin{center}
\includegraphics[angle=0, totalheight=7cm]{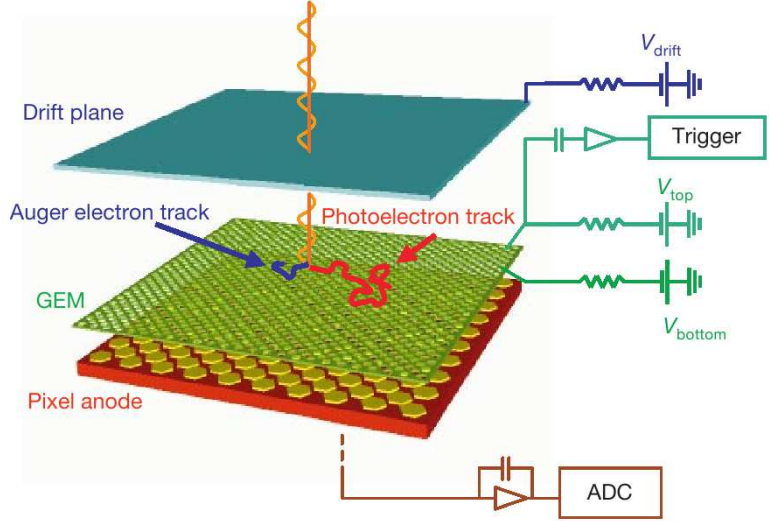}
\end{center}
\caption{\small Principle of operation of the GPD.} 
\label{fig:GPD}
\end{figure}

The Gas Pixel Detector\cite{Costa2001, Bellazzini2006, Bellazzini2007}, developed by the INFN of 
Pisa, is one of the most advanced X-ray polarimeter based on the photoelectric effect. It is
basically a gas detector with fine 2-D position resolution, made of a gas cell with a beryllium
X-ray window, a gas electron multiplier (GEM) and, below it, of a pixellated plane which collects
the charges generated in the gas. Its operating principle is shown in Fig.~\ref{fig:GPD}. When a
photon crosses the thin beryllium window and is absorbed in the gas, a photoelectron is emitted
preferentially along the electric field of the absorbed photon, i.e. in the direction of
polarization. As the photoelectron propagates, it is scattered by charges in the nuclei and loses
its energy by ionization. Its path is traced by the generated electron-ion pairs, which are
amplified by the GEM and collected on the fine sub-divided pixel detector. Hence the detector sees
the projection of the track of the photoelectron.

When the incident radiation is linearly polarized, the histogram of the angles of emission of a
large number of photoelectrons is modulated with an amplitude proportional to the degree of
polarization. The Gas Pixel Detector derives the amplitude and the phase of this modulation
by analyzing the projection of the track on the detector and then reconstructing the angle of
emission of each photoelectron.

This approach is successful only if the tracks are resolved by the detector in many pixels.
Indeed, photoelectrons are subjected to nuclear scatterings in the gas and then the first
part of the track must be resolved, since it's the only part that carries memory of the
polarization. Moreover, the ejection of the photoelectron is very likely followed by the production
of an Auger electron, which must be distinguished because its direction of emission is isotropic.

The current version of the GPD has 105k pixels with 50~$\mu$m pitch in a hexagonal pattern. The
version employed for the measurements reported below was filled and sealed with a mixture
composed by 30\% helium and 70\% DME at 1~atm, with a gas cell 1~cm thick and the GEM at 400~$\mu$m
from the detector. Since the helium is nearly transparent for photons at a few keV, the DME acts as
quencher and absorber of photons.

\section{Analysis of the photoelectric differential cross section} \label{sec:PDC}

\begin{figure}[htbp]
\begin{center}
\includegraphics[angle=0,totalheight=8cm]{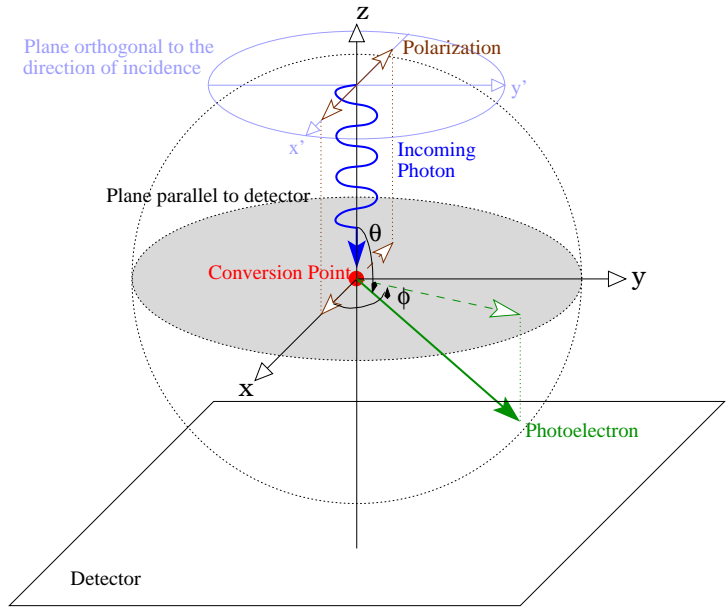}
\end{center}
\caption{\small Definition of the emission angle of photoelectrons.}
\label{fig:EmissionAngles}
\end{figure}

The direction of emission of a photoelectron carries memory of the polarization of the absorbed
photon, namely the probability that a photoelectron is emitted in the direction of polarization is
modulated with a $\cos^2\phi$ term, where $\phi$ is the angle between the direction of emission and
polarization (see Fig.~\ref{fig:EmissionAngles}).

If we consider only the absorption of shells with spherical symmetry, i.e. 1s and 2s, which give the
larger contribution to the photoelectric absorption in the working condition of the GPD, the
differential cross section of the photoelectric effect is:
\begin{equation}
\frac{d\sigma}{d\Omega} \propto \frac{\sin^2\theta
\cos^2\phi}{\left(1+\beta\cos\theta\right)^4},\label{eq:dSdO}
\end{equation}
where $\theta$ is the angle between the direction of the photon and that of the photoelectron (see
again Fig.~\ref{fig:EmissionAngles}). 

\begin{figure}[htbp]
\begin{center}
\includegraphics[angle=0,totalheight=9cm]{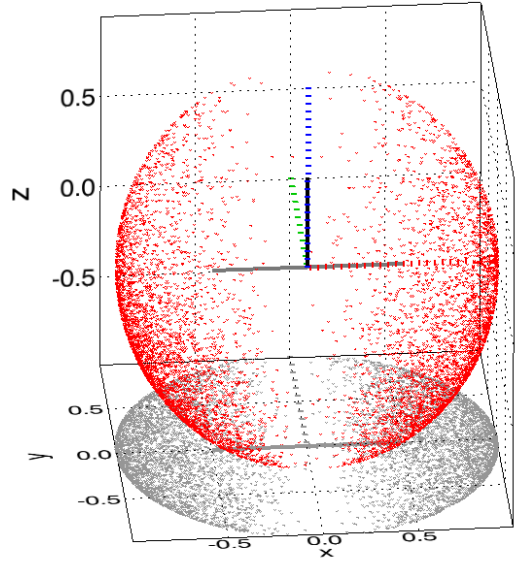}
\end{center}
\caption{\small Forward folding due to the relativistic correction to the differential
photoelectric cross section. The direction of incidence of the photons is the black continuous line,
while the direction of polarization is gray. The photons are absorbed in the center of the sphere of
unit radius, in the point (0,0,0), and the directions of emission of photoelectrons are traced with
the red points on the sphere. The projection of the plot on the \emph{xy} plane is also
shown. The red, green and blue dashed lines are respectively the \emph{x}, \emph{y} and
\emph{z}~axes. The energy of the photoelectrons was set to 50~keV to show the relativistic effects
more clearly.}
\label{fig:RelCor_Orthogonal}
\end{figure}

The Eq.~\ref{eq:dSdO} includes the relativistic correction
$\frac{1}{\left(1+\beta\cos\theta\right)^4}$, where $\beta$ is the velocity of the
emitted electron in units of the speed of light. Without this relativistic correction, the
photoelectrons are emitted preferentially into the plane perpendicular to the direction of
incidence, with a symmetry between the emission above and below this plane. The lack of this
symmetry induced by the relativistic correction makes the emission probability higher in the
semi-space opposite with respect to the direction of photon. This effect of ``forward folding'' is
shown in Fig.~\ref{fig:RelCor_Orthogonal}, where the energy of photoelectrons is 50~keV to stress
the relativistic effects. The photons are absorbed in the center of the sphere of
unit radius, in the point (0,0,0), and the directions of emission of photoelectrons are traced by
the red points which are concentrated in the semi-space \emph{z}$<0$. Note also that the projection
of the directions of emission on the \emph{xy} plane is not changed by the relativistic effects,
namely it still follows a $\cos^2\phi$ dependency and no photoelectrons are emitted
orthogonally to the direction of polarization. 

The modulation of the direction of emission of photoelectrons, as measured by a
polarimeter based on the photoelectric effect like the GPD, can be derived from Eq.~\ref{eq:dSdO}.
At this aim, we must distinguish between the intrinsic modulation $\mathcal{M}_0'(\phi')$, namely
the modulation in the reference frame \emph{x'y'z'} of the photons (see
Fig.~\ref{fig:EmissionAngles} and Fig.~\ref{fig:IncidenceAngles}), the intrinsic one in the
reference of the detector \emph{xyz},
$\mathcal{M}_0(\phi)$, and that measured by a polarimeter
$\mathcal{M}(\phi)$. 

If the hypothesis of absorption from only shells with spherical symmetry is
satisfied, the modulation $\mathcal{M}_0'$ for polarized photons can be simply derived from
Eq.~\ref{eq:dSdO} by integrating over angles $\theta$:
\begin{equation}
^{pol}\mathcal{M}_0'(\phi') = \frac{d \sigma'}{d\phi'} \propto \int_0^\pi \frac{\sin^2\theta'
\cos^2\phi'}{\left(1+\beta\cos\theta'\right)^4} \sin\theta' d\theta' \propto \cos^2\phi',
\end{equation}
where the term $\sin\theta'$ derives from $d\Omega' = \sin\theta' d\theta' d\phi'$

In the same way, if photons are unpolarized, 
\begin{equation}
^{unpol}\mathcal{M}_0'(\phi') = \frac{d \sigma'}{d\phi'} \propto \int_0^\pi
\frac{\sin^2\theta'}{\left(1+\beta\cos\theta'\right)^4} \sin\theta' d\theta' \propto 1.
\end{equation}

If the photons are incident orthogonally to the detector, the frames of reference \emph{x'y'z'}
and \emph{xyz} are identical (see Fig.~\ref{fig:EmissionAngles}), and hence:
\begin{equation}
\mathcal{M}_0(\phi) \equiv \mathcal{M}_0'(\phi').
\end{equation}

In the hypothesis that the polarimeter has negligible systematic effects, the direction of the
emission will be reconstructed correctly with a probability, which in general depends on energy,
identical in every direction. Hence, the modulation $\mathcal{M}(\phi)$ measured by the detector is
proportional to the intrinsic one (in the frame of reference of the detector), plus a constant which
takes into account the errors in the process of measurement, i.e. the probability to reconstruct
the direction of emission incorrectly:
\begin{equation}
\mathcal{M}(\phi) = M\cdot\mathcal{M}_0(\phi) + C \label{eq:M} ~~~\mbox{with}~~~\left\{
\begin{array}{rcl}
^{pol}{M}_0(\phi) &=& \cos^2\phi \\
^{unpol}{M}_0(\phi) &=& 1
\end{array}
\right.,
\end{equation}
where $M$ and $C$ are constants. Eq.~\ref{eq:M} is the canonic function employed to fit the
modulation of the histogram of the direction of photoelectrons with an X-ray polarimeter like the
GPD.

The modulation factor $\mu$, namely the amplitude of the response of the instrument for completely
polarized photons, can be derived
from Eq.~\ref{eq:M}:
\begin{equation}
\mu =
\frac{\mbox{max}\left\{\mathcal{M}\right\}-\mbox{min}\left\{\mathcal{M}\right\}}{\mbox{max}\left\{
\mathcal{M}\right\}+\mbox{min}\left\{\mathcal{M}\right\}} = \frac{M}{M+2\cdot C},
\end{equation}
where the last passage is allowed only if the $\mathcal{M}_0(\phi)$ function is normalized to
unity.


\section{Expected modulation for inclined photons}
\label{sec:InclinedEffetcts}

\begin{figure}[htbp]
\begin{center}
\subfigure[\label{fig:FocusingOptics}]{\includegraphics[angle=0,totalheight=9cm]{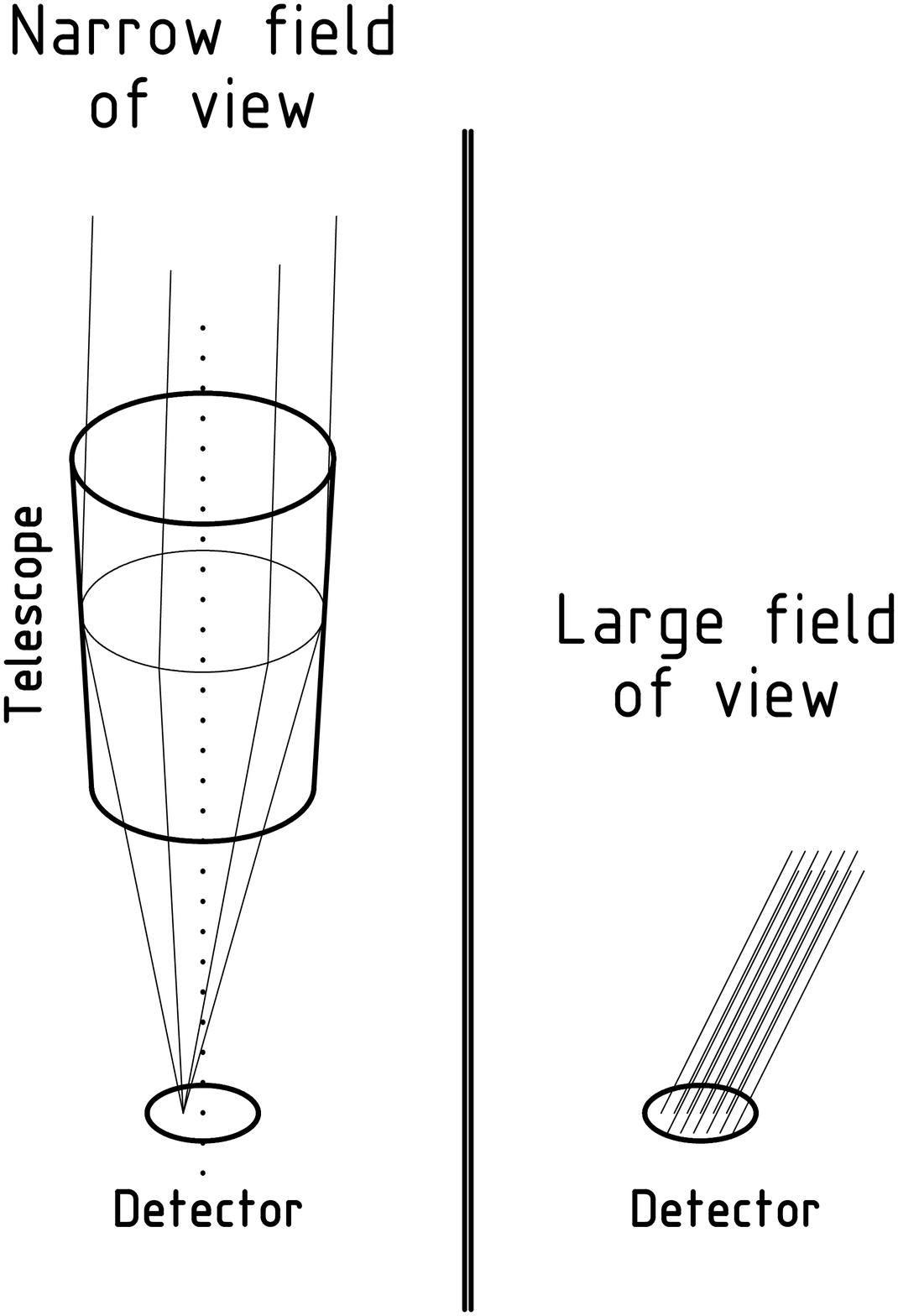}
}
\subfigure[\label{fig:IncidenceAngles}]{\includegraphics[angle=0,
totalheight=7cm]{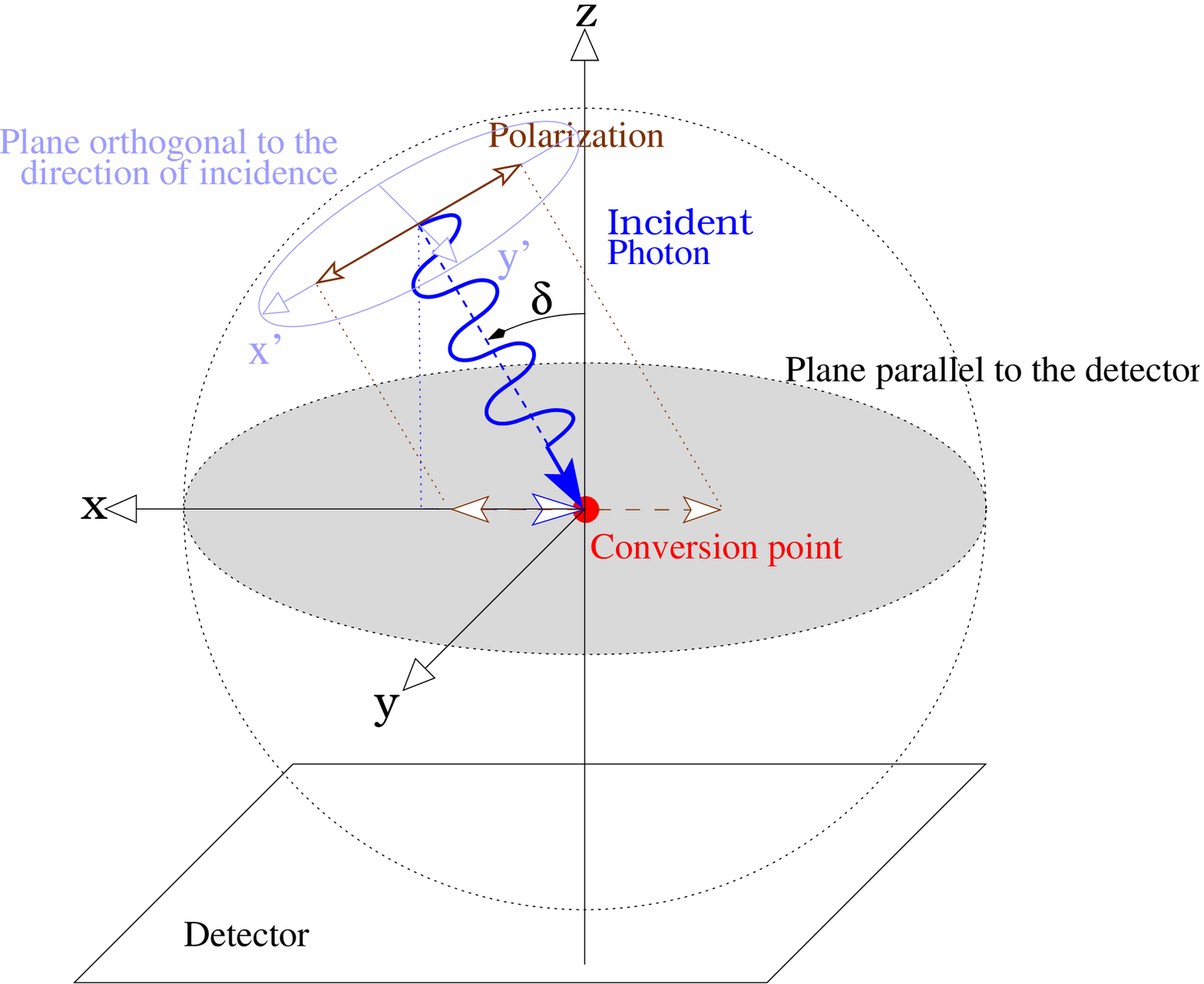}}
\end{center}
\caption{\small ({\bf a}) Differences between the narrow and the large field of view instruments.
({\bf b}) Angles which define the direction of incidence of photons.}
\end{figure}

If the GPD is employed as narrow field instrument with an X-ray optics, the photons are incident
nearly perpendicularly to the plane of detector, with an angle determinated by the focusing of the
optics but always of the order of a few degrees. In this case, systematic effects are small and also
reduced, even out of the optical axis, since the azimuthal angles are different for each photon and
hence different systematic effects sum incoherently (see Fig.~\ref{fig:FocusingOptics}). Instead,
for instruments with large field of view, photons coming from a source are all incident with a large
and identical inclination.

Here we want to explore the systematic effects arising when the inclination angle $\delta$ is
greater than a few degrees, for polarized and unpolarized radiation. We limit ourself to the
condition of inclination orthogonal to the direction of polarization with the following assumptions:
\begin{enumerate}
\item the direction of polarization is along the \emph{x} axis;
\item the inclination of an angle $\delta$ is performed around \emph{y} axis (see
Fig.~\ref{fig:IncidenceAngles}).
\end{enumerate}

The function $\mathcal{M}_0(\phi)$ can be derived with the same procedure presented in
Sec.~\ref{sec:PDC}, with the important difference that, in this case, the references \emph{x'y'z'}
and \emph{xyz} are not identical. We must first transform the functions:
\begin{equation}
\begin{array}{rcl}
\frac{d \sigma'}{d\Omega'}\mid_{pol} &\propto& \frac{\sin^2\theta'
\cos^2\phi'}{\left(1+\beta\cos\theta'\right)^4}\\
\frac{d \sigma'}{d\Omega'}\mid_{unpol} &\propto&
\frac{\sin^2\theta'}{\left(1+\beta\cos\theta'\right)^4}
\end{array},
\end{equation}
from the \emph{x'y'z'} reference to the \emph{xyz} one and then perform the integration over
$\theta$. Since the relativistic correction is not easily integrable, we approximated this term at
its first order:
\begin{equation}
\frac{1}{\left(1+\beta\cos\theta\right)^4} \approx 1-4\beta\cos\theta.
\end{equation}

Eventually, we obtain:
\begin{equation}
\begin{array}{rcl}
^{pol}\mathcal{M}_0(\phi,\delta,\beta) &=& 
\left(-\frac{3}{2}\pi\beta\sin\delta\cos^2\delta\right)\cos^3\phi + \left(\frac{4}{3}
\cos^2\delta\right)\cos^2\phi
+\left[\frac{\pi}{2}\beta\left(2\sin\delta-3\sin^3\delta\right)\right]\cos\phi+\\
&&+\left(\frac{2}{3}\sin^2\delta\right);\\
\\
^{unpol}\mathcal{M}_0(\phi,\delta,\beta) &=& 
\left(\frac{3}{2}\pi\beta\sin^3\delta\right)\cos^3\phi + \left(-\frac{4}{3}
\sin^2\delta\right)\cos^2\phi
+\left[\pi\beta\left(\frac{3}{2}\sin\delta\cos^2\delta-2\sin\delta\right)\right]\cos\phi+\\
&&+\left(2-\frac{2}{3}\cos^2\delta\right).
\end{array}
\label{eq:Functions}
\end{equation}

Note that the functions $^{pol}\mathcal{M}_0(\phi,\delta,\beta)$ and
$^{unpol}\mathcal{M}_0(\phi,\delta,\beta)$ are quite complex, but depend only on the parameters
$\beta$, $\delta$, namely on the velocity and on the inclination of the photons. Note also
that, when $\delta=0^\circ$, the functions are proportional to the functions reported in
Eq.~\ref{eq:M}, obtained when photons are incident perpendicularly to the detector.

The function $^{pol}\mathcal{M}_0(\phi,\delta,\beta)$ expresses the presence of a intrinsic
constant term in the modulation curve for completely polarized and inclined photons, namely a
costant contribution not due to the errors of measurement:
\begin{equation}
^{pol}\mathcal{M}_0(\phi=90^\circ,\delta,\beta)=\frac{2}{3}\sin^2\delta\neq0~~~\mbox{if}
~~~\delta\neq0.
\end{equation}

Moreover, when $\beta\neq0$ and hence the relativistic correction is introduced, there is an
asymmetry between the two peaks corresponding to the direction of polarization, which instead are
identical when photons are orthogonal to the detector, deriving from the terms proportional to
$\cos^3\phi$ and $\cos\phi$.

\begin{figure}[htbp]
\begin{center}
\includegraphics[angle=0,totalheight=9cm]{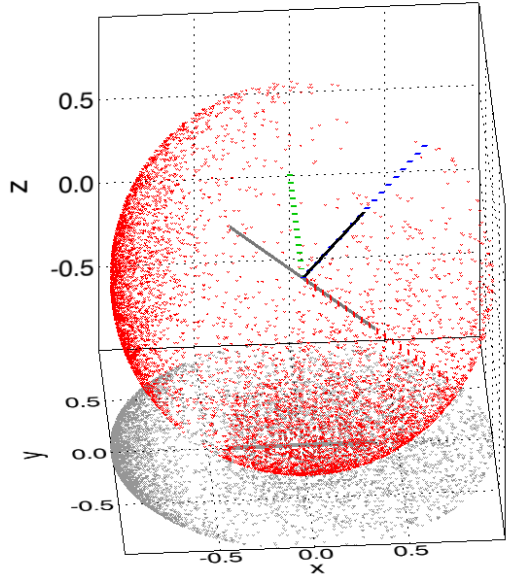}
\end{center}
\caption{\small The same as Fig.~\ref{fig:RelCor_Orthogonal}, but for photons which are inclined
of 30~degrees. Note that in this case there are tracks of photoelectrons which are projected along
the \emph{y} axis, namely orthogonally to the direction of polarization. Note also the different
intensities of projections on the left and on the right, which indicate a different probability of
emission.}
\label{fig:RelCor_Inclined}
\end{figure}

Inclined systematic effects can be visualized with the projection of the directions of emission of
photoelectrons in the \emph{xy} plane. While for orthogonal photons no photoelectrons tracks are
projected in the \emph{y} direction, namely in the direction orthogonal to the polarization (see
Fig.~\ref{fig:RelCor_Orthogonal}), in case of inclined photons there is emission even in this
direction (see Fig.~\ref{fig:RelCor_Inclined}). Moreover, the emission on the left is more probable
than that on the right (see again Fig.~\ref{fig:RelCor_Inclined}) and this induces the asymmetry
between the peaks of the modulation cuve.

In the case of unpolarized photons, a spurious modulation appears. In absence of the relativistic
correction, this contribution would be of the same form as a polarized signal:
\begin{equation}
^{unpol}\mathcal{M}_0(\phi,\delta,\beta=0) =
\left(-\frac{4}{3}\sin^2\delta\right)\cos^2\phi+\left(2-\frac{2}{3}\cos^2\delta\right).
\end{equation}

However the presence of the relativistic correction introduces an asymmetry that allows, at least in
principle, to distinguish it from the signal of polarized radiation.

%

\section{Measurement of the systematic effects for inclined radiation} \label{sec:Measurements}

\begin{figure}[htbp]
\begin{center}
\includegraphics[angle=0,totalheight=9cm]{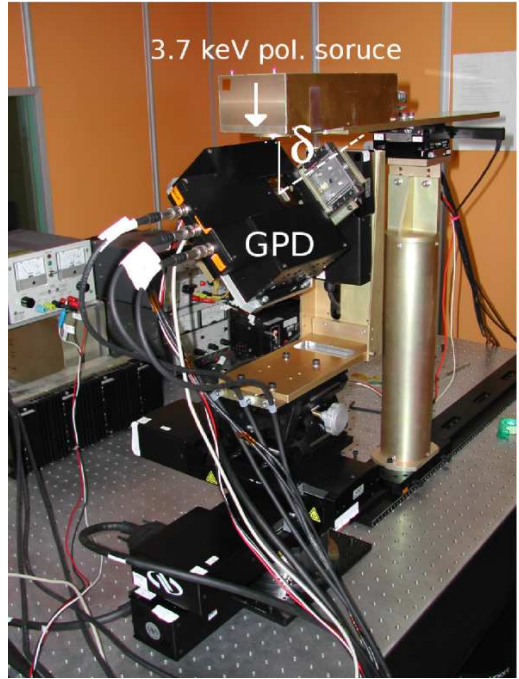}
\end{center}
\caption{\small The setup employed to perform the measurements at large angle of inclination. A
complete description of the facility used is presented by Muleri et al.\cite{Muleri2008b}.}
\label{fig:Setup}
\end{figure}

Muleri et al.\cite{Muleri2008b} presented a facility for the calibration of the next generation
X-ray polarimeters, which allows to study the response of detectors to inclined, polarized and
unpolarized radiation. We then employed this facility to measure the systematic effects presented
in Sec.~\ref{sec:InclinedEffetcts} (see Fig.~\ref{fig:Setup}).

We employed the 105k pixels, 50$\mu$m pitch, sealed version of the Gas Pixel Detector, described by
Bellazzini et al.\cite{Bellazzini2007}, and filled with 30\% helium and 70\% DME mixture. Data were
analyzed in the standard way, without any data selection in the shape of the tracks.

\begin{figure}[htbp]
\begin{center}
\subfigure[\label{fig:426_cos2}]{\includegraphics[angle=0,width=5.6cm]{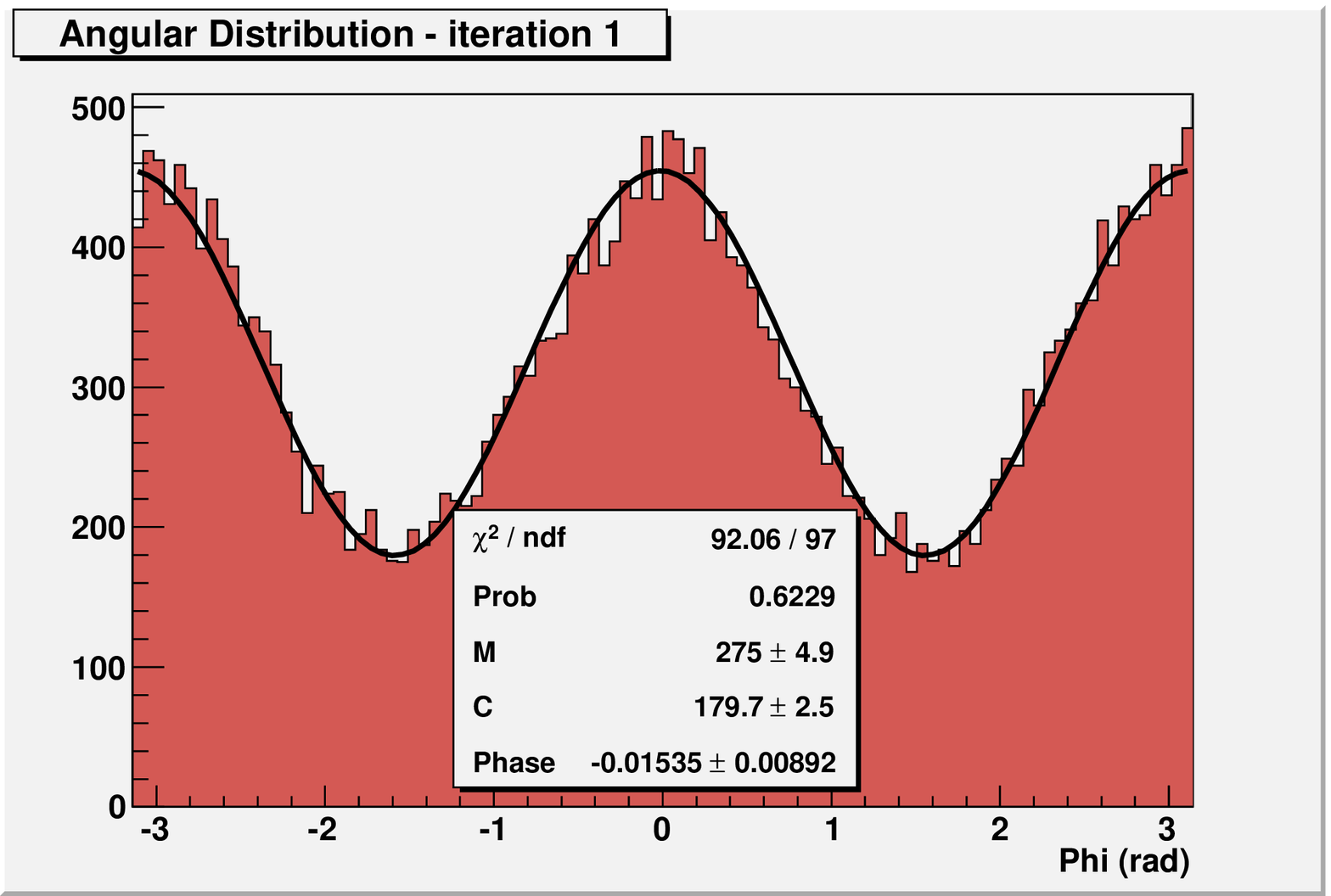}}
\subfigure[\label{fig:429_cos2}]{\includegraphics[angle=0,width=5.6cm]{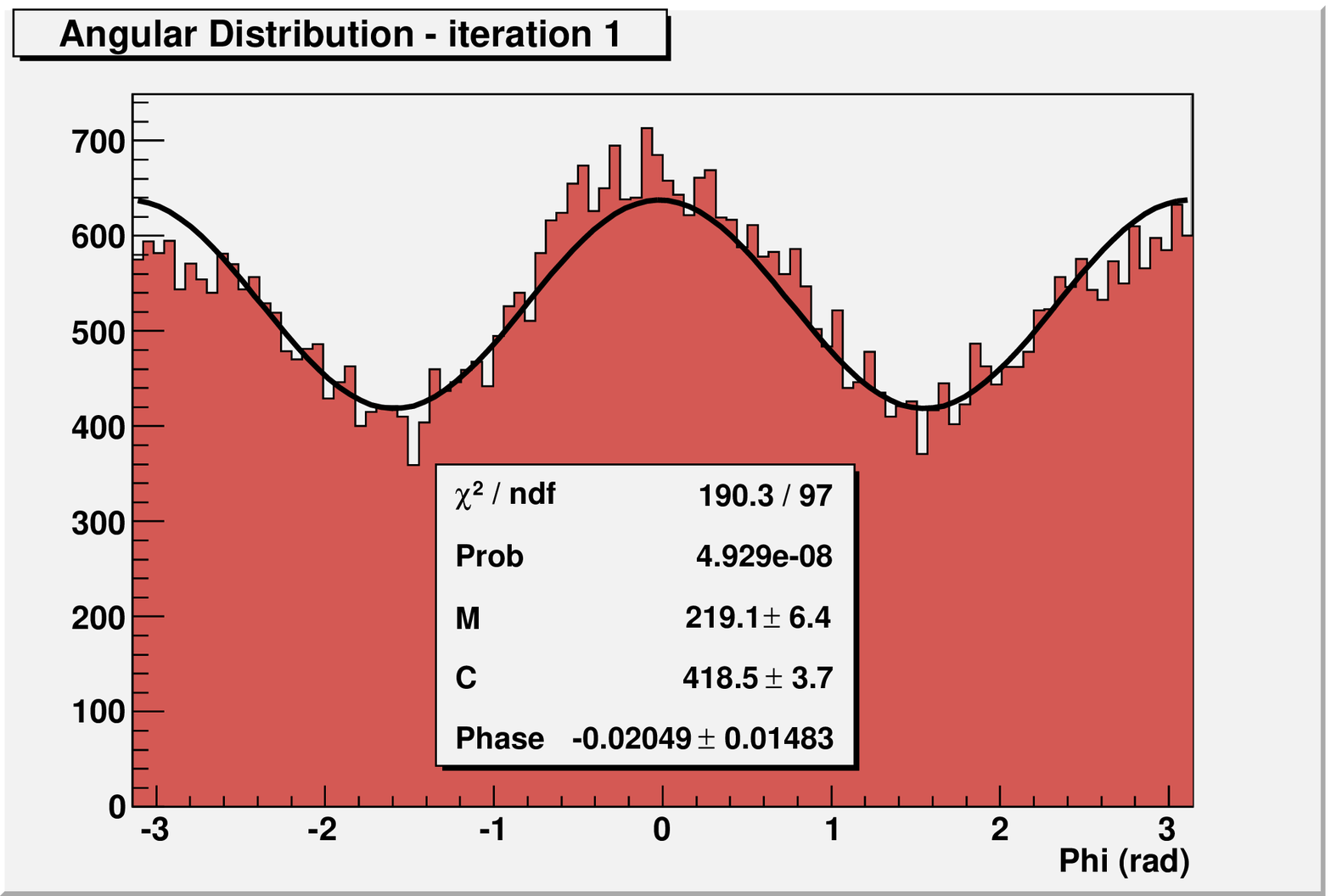}}
\subfigure[\label{fig:542_cose}]{\includegraphics[angle=0,width=5.6cm]{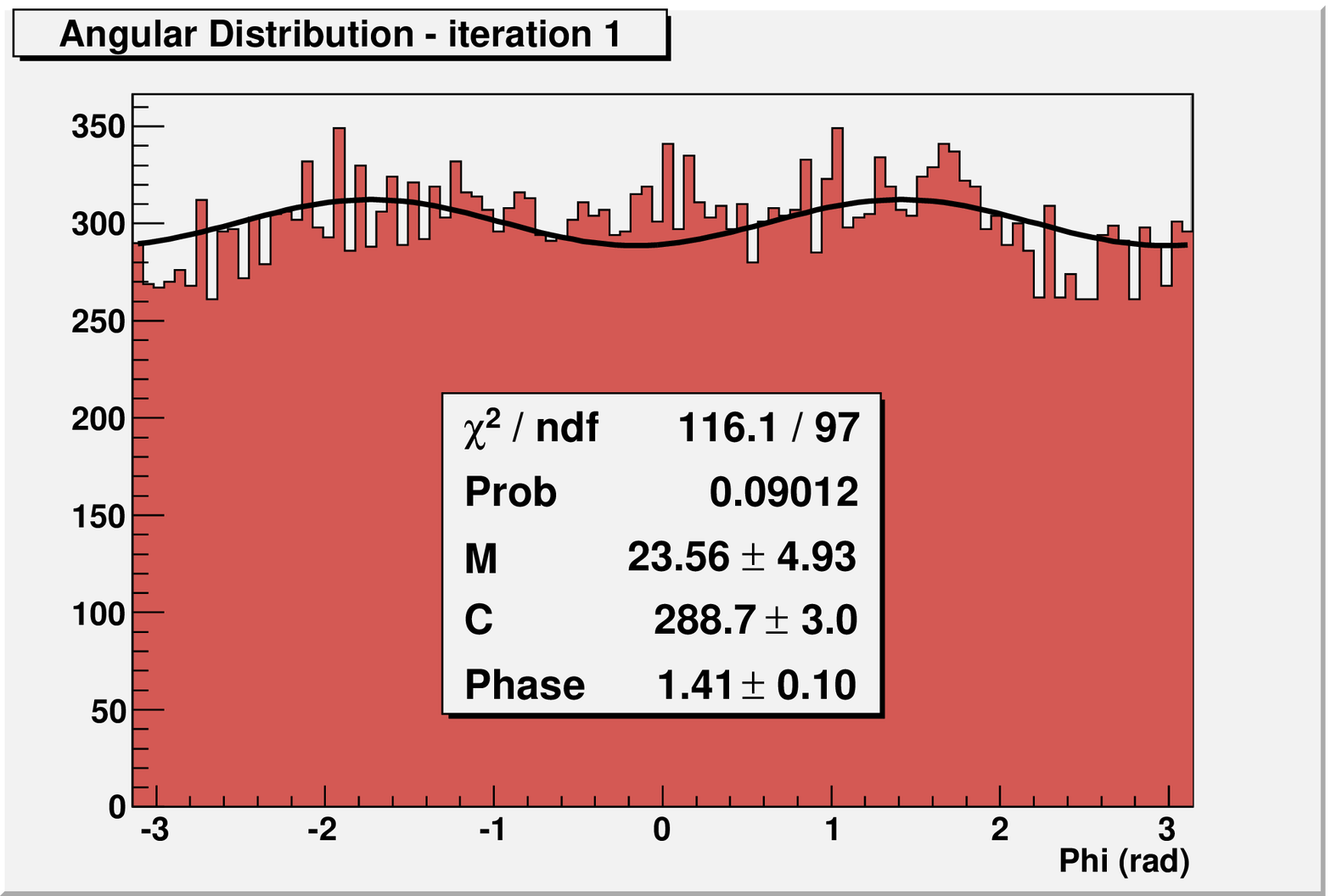}}
\end{center}
\caption{\small Response of the GPD to inclined, polarized and unpolarized radiation, fitted with a
standard $\cos^2$ function. ({\bf a}) Modulation measured when 3.7~keV polarized photons, produced
by means of Bragg diffraction at nearly 45~degrees, are incident orthogonally to the detector. ({\bf
b}) Response to the same kind of radiation as ({\bf a}) when the beam is inclined 40~degrees with
respect to the orthogonal of the detector. ({\bf c}) Modulation measured when 5.9~keV unpolarized
photons are incident with an inclination of 30~degrees.}
\label{fig:cos2}
\end{figure}

In Fig.~\ref{fig:cos2} we report the modulation measured with the GPD in three different
situations, fitted with the standard $\cos^2$ function:
\begin{equation}
\mathcal{M}(\phi) = M\cdot\cos^2(\phi-\phi_0) + C.
\end{equation}

In particular, in Fig.~\ref{fig:426_cos2} we show the response of the instrument to 3.7~keV
polarized photons which are incident orthogonally to the detector. The fit is very good, as
reported in Table~\ref{tab:Fits}, and a modulation factor of about 43\% is obtained. Instead in
Fig.~\ref{fig:429_cos2} we report the modulation when the same kind of photons, i.e. 3.7~keV
polarized ones, are incident with an inclination of 40~degrees. In this case the fit is
extremely poor, the peaks are asymmetric and the modulation factor is by far lower, about
21\%. These systematic effects were those expected from the analysis performed in the
Sec.~\ref{sec:InclinedEffetcts}, i.e. a reduction of the modulation due to a intrinsic constant
term and the asymmetry due to the relativistic correction. Eventually in Fig.~\ref{fig:542_cose} we
report the response of the instrument to unpolarized 5.9~keV obtained from a Fe$^{55}$
radioactive source, inclined with respect to the perpendicular of the detector of 30~degrees. As
expected, in this case a spurious and asymmetric modulation is present and it mimics the
effect of a polarized beam at the level of a few percent.

\begin{table}[htbp]
\begin{center}
\begin{tabular}{c|c|c|c|c|c|c|c} 
Polarization & Energy (keV) & Inclination ($^\circ$)& function & M & C & $\mu$ & $\chi^2_{reduced}$
\\
\hline
\hline
Pol. & 3.7 & $0^\circ$ & $\cos^2$ & 275.0$\pm$4.9 & 179.7$\pm$2.5 & 0.4335 & 0.95\\
Pol. & 3.7 & $40^\circ$ & $\cos^2$ & 219.1$\pm$6.4 & 418.5$\pm$3.7 & 0.2075 & 1.96\\
Unpol. & 5.9 & $30^\circ$ & $\cos^2$ & 23.56$\pm$4.93 & 288.7$\pm$3.0 & --- & 1.20\\
\hline
Pol. & 3.7 & $0^\circ$ & $^{pol}\mathcal{M}$ & 275.0$\pm$4.9 & 179.7$\pm$2.5 & 0.4335 & 0.95\\
Pol. & 3.7 & $40^\circ$ & $^{pol}\mathcal{M}$ & 334.0$\pm$9.5 & 338.5$\pm$5.6 & 0.3304 & 1.24\\
Unpol. & 5.9 & $30^\circ$ & $^{unpol}\mathcal{M}$ & 78.$\pm$11. & 196$\pm$15 & --- & 0.93\\
\end{tabular}
\caption{Parameters of the fits shown in the Fig.~\ref{fig:cos2} and in the Fig.~\ref{fig:f1g1}.}
\label{tab:Fits}
\end{center}
\end{table}

\begin{figure}[htbp]
\begin{center}
\subfigure[\label{fig:426_f1}]{\includegraphics[angle=0,width=5.6cm]{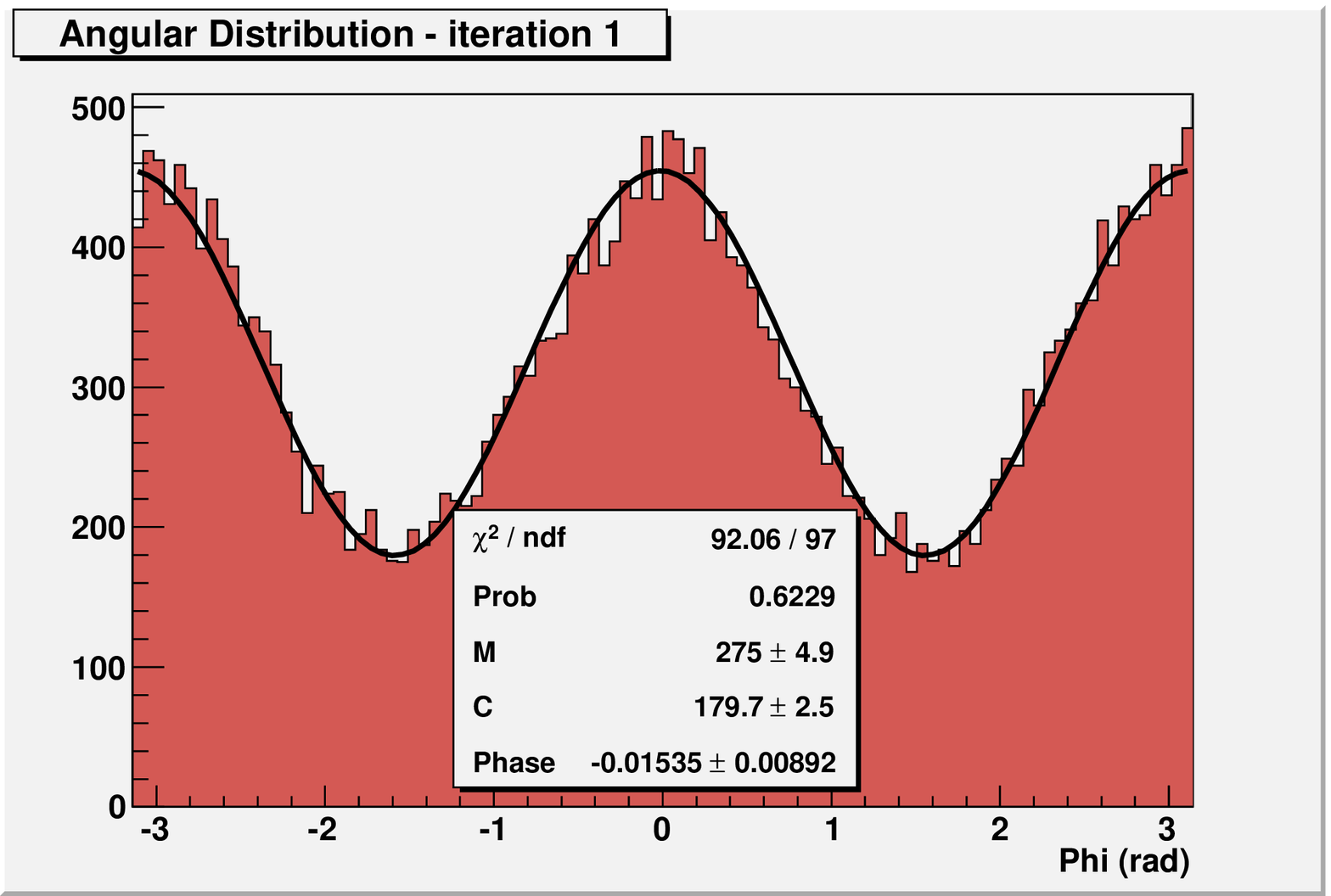}}
\subfigure[\label{fig:429_f1}]{\includegraphics[angle=0,width=5.6cm]{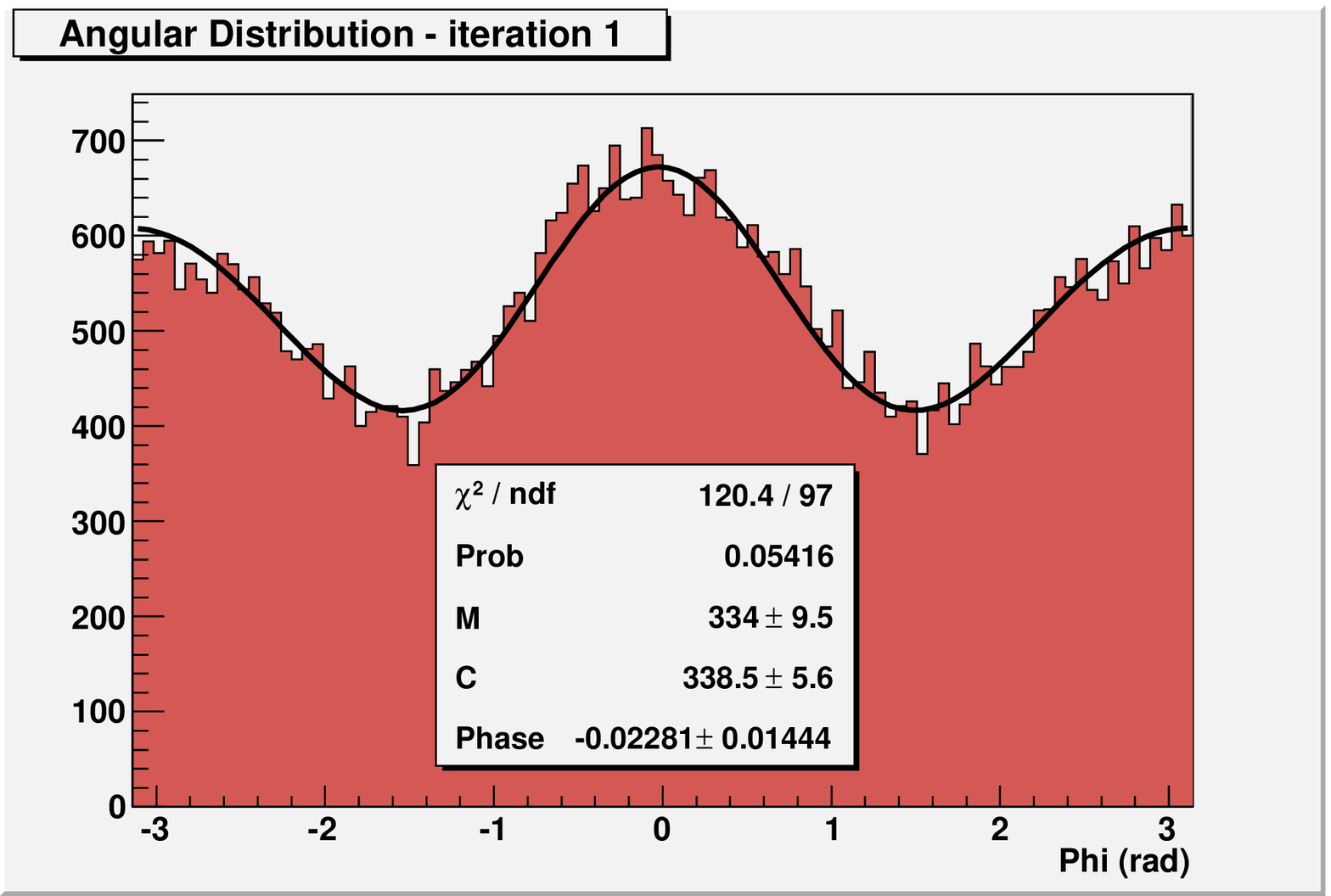}}
\subfigure[\label{fig:542_g1}]{\includegraphics[angle=0,width=5.6cm]{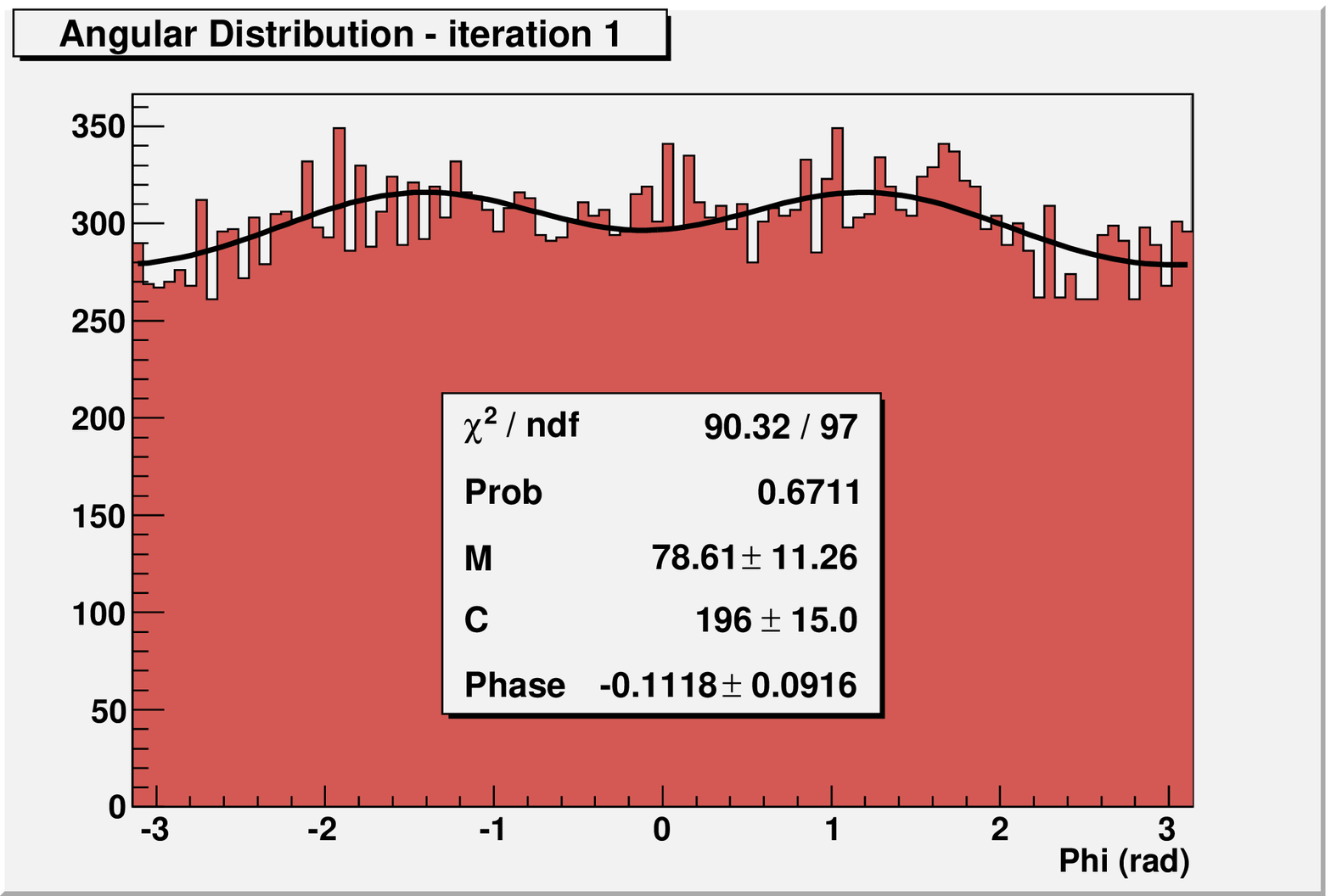}}
\end{center}
\caption{\small The same as Fig.~\ref{fig:cos2}, but in this case the fits are perfromed with the
functions derived in Sec.~\ref{sec:InclinedEffetcts}. The parameters $\beta$ and $\delta$ are fixed
at their expected values. The results of the fits are reported in Table~\ref{tab:Fits}.}
\label{fig:f1g1}
\end{figure}

In Fig.~\ref{fig:f1g1} we report the same data presented in Fig.~\ref{fig:cos2} but fitted
with the functions developed in Sec.~\ref{sec:InclinedEffetcts} (Eq.~\ref{eq:Functions}).
Since we want to verify the systematic effects derived in Sec.~\ref{sec:InclinedEffetcts}, we
assumed that the inclination and the velocity of the photoelectrons were known. Hence, we assumed
that the shape of the modulation is fixed, and only the normalization is allowed to change. Then, in
the case of polarized photons, we employed:
\begin{equation}
^{pol}\mathcal{M}(\phi,\phi_0,\delta=\overline{\delta},\beta=\overline{\beta}) =
M\cdot\left\{\frac{1}{\frac{4}{3}-\frac{2}{3}\sin^2\overline{\delta}+\frac{\pi\overline{\beta}}{2}
\left|\sin\overline{\delta}\right|}
\left [^{ pol} \mathcal{M}_0(\phi-\phi_0 ,\overline{\delta},\overline{\beta})\right]\right\} +C,
\label{eq:M_pol}
\end{equation}
where $\overline{\delta}$ and $\overline{\beta}$ are constants. As in the case of Eq.~\ref{eq:M},
we fitted data with a function which is proportional to the intrinsic modulation, plus a constant
which takes into account the error in the process of measurement. We introduced the phase $\phi_0$
and the factor
$\frac{1}{\frac{4}{3}-\frac{2}{3}\sin^2\delta+\frac{\pi\beta}{2}\left|\sin\delta\right|}$ to
normalized to unity the function $^{pol}\mathcal{M}_0(\phi,\delta,\beta)$ so that we can calculate
the modulation factor with the usual formula $\mu = \frac{M}{M+2\cdot C}$.

Instead, we employed:
\begin{equation}
^{unpol}\mathcal{M}(\phi,\phi_0,\delta=\overline{\delta},\beta=\overline{\beta}) =
M\cdot\left\{^{unpol}\mathcal{M}_0(\phi-\phi_0 , \overline{\delta} , \overline{\beta})\right\} +C
\end{equation}
for fitting data for unpolarized radiation. Even in this case the shape of the modulation is fixed
and only the normalization can change.

The results of the fits are reported in Table~\ref{tab:Fits}. The agreement is quite good for all
the measurements done. In particular, taking into account the systematics, we can recover a good
modulation factor even at 40~degrees, about 33\% against the 43\% on axis. Instead the fit
with the $\cos^2$ gives a modulation of about 21\%, since it doesn't take into account that there is
a intrinsically unmodulated contribution. 

Eventually we want also stress that the systematic effects presented are only visible in the
unfolded modulation curve, namely if the directions of photoelectrons are reconstructed on the
whole round angle. Indeed, by assuming that the modulation curve is periodic on 180~degrees, these
systematics are confused and often not distinguishable from the actual signal.

\section{Conclusion}

We have presented a feasibility study for testing the possible use of the Gas Pixel Detector as an
X-ray photoelectric polarimeter with large field of view. Accurately analyzing the differential
cross section of the photoelectric absorption and performing measurements with polarized and
unpolarized inclined photons, we have found that systematic effects on the measured modulation
exist for both polarized and unpolarized radiation when photons are not incident orthogonally to the
detector. For polarized radiation, these systematics induce an intrinsic unmodulated contribution
and an asymmetry between the peaks of the modulation curve corresponding to the direction of
polarization. While the former effect arises from the inclination of the photons and tends to reduce
the modulation factor, the latter is due to the relativistic correction of the differential
photoelectric cross section which can only be distinguished in the unfolded modulation curve.
Instead the systematics in the case of unpolarized radiation mimic the modulation induced by
polarized photons and only a relativistic asymmetry, again visible only in the unfolded modulation
curve, can, in principle, disentangle the two situations.

After discussing the systematics arising when polarized or unpolarized inclined beams are
incident on the detector, we have also presented functions to fit the modulation curves and correct
the systematic effects. We have only restricted the analysis in one of the simplest situation,
assuming that the direction of polarization is parallel to the axis of inclination. Even if we
can't today establish the feasibility of photoelectric polarmeter with large field of view, good
results are obtained by assuming that the inclination and the velocity of the photoelectrons are
known. This corresponds to assume that the instrument has both good imaging and spectroscopic
capabilities, that is the case of the GPD. 


\section*{Acknowledgments}

FM acknowledges financial support from Agenzia Spaziale Italiana (ASI) under contract
ASI~I/088/06/0.

\bibliography{References}   
\bibliographystyle{spiebib}   

\end{document}